\documentclass[11pt]{article}

\usepackage[utf8]{inputenc}
\usepackage[T1]{fontenc}
\usepackage{fullpage}
\usepackage{url}
\usepackage{amsmath}
\usepackage{amssymb}
\usepackage{amsfonts}
\usepackage{amsthm}
\usepackage{enumerate}
\usepackage{epsfig}
\usepackage{color}
\usepackage{dsfont}
\usepackage{pdfpages}
\usepackage{graphicx}
\usepackage{calligra}
\usepackage{subcaption}
\usepackage{float}

\newcommand{\mc}{\mathcal}

\newcommand{\bec}{\begin{center}}
\newcommand{\eec}{\end{center}}
\newcommand{\bi}{\begin{itemize}}
\newcommand{\ei}{\end{itemize}}
\newcommand{\btab}{\begin{tabular}}
\newcommand{\etab}{\end{tabular}}

\newcommand{\begit}{\begin{enumerate}[$\bullet$]}
\newcommand{\eeit}{\end{enumerate}}

\newcommand{\dd}{\textrm{d}}

\newcommand{\proba}{\mathcal{P}}

\newcommand{\pproba}{P}
\newcommand{\pathInt}{\mathcal{D}}

\newcommand{\pathGain}{\mathcal{G}}
\newcommand{\pathRisk}{\mathcal{R}}
\newcommand{\pathCost}{\mathcal{C}}

\newcommand{\pathCostP}{\mathcal{P}}

\newcommand{\lowerBound}{\ell}
\newcommand{\upperBound}{{\calligra u}}

\newcommand{\lambdaHalf}{\frac{1}{2}}
\newcommand{\approxim}{\sim}
\newcommand{\condition}{\left\vert\vphantom{\frac{0}{0}}\right.}

\title{Equations and Shape of the Optimal Band Strategy}
\author{Joachim de Lataillade, Ayman Chaouki\\
\\
Capital Fund Management\\
  23 rue de l'Université, 75007 Paris, France}
\date{\today}

\begin{document}

\newtheorem{Proposition}{Proposition}
\newtheorem{Theorem}{Theorem}

\maketitle

\begin{abstract}
  We consider the problem of the optimal trading strategy in the
  presence of a price predictor, linear trading costs and a quadratic
  risk control. The solution is known to be a band system, a policy
  that induces a no-trading zone in the positions space.
  Using a path-integral method introduced in a previous
  work, we give equations for the upper and lower edges of this band,
  and solve them explicitly in the case of an Ornstein-Uhlenbeck predictor. We then explore the shape of this solution and derive its asymptotic behavior for large values of the predictor,
  without requiring trading costs to be small.
 \end{abstract}


\section{Introduction}

Price returns on financial markets are by nature very difficult to
predict, and the goal of statistical arbitrage is to find small but
significant predictive patterns in all available data. However, from a
practitioner's perspective, the prediction of the price is only an
ingredient in the building of a trading system: controlling the risk
taken by this system, and avoiding high costs when trading, are
crucial elements of success.

In the present paper, we focus on the optimisation of trading in a
specific case: we consider the single-asset case, where the risk is
controlled through a penalty on the square of the exposure (or
position) on that asset, and with a linear cost of trading of the form
$\Gamma|Q|$, where $Q$ is the quantity bought or sold at a given time.
Because of the relation between costs and market impact
models~\cite{toth, minimalModelImpact}, quadratic or at least superlinear models of costs
are often considered~\cite{strategiesImpact, quadCosts}. Linear transaction costs are
nonetheless relevant when considering market and brokerage fees, or
costs for crossing the bid-ask spreads, as they become dominant for
small trading amounts.

Systems with linear (aka. proportional) transaction costs have been considered on many
occasions in the literature~\cite{davisNorman, shreveSoner, cmeCosts,
  meanReversion, martinMulti}, with a focus on particular on the limit
of small transaction costs~\cite{johannesReview, johannesLinAndQuad, cfmLinAndQuad}. The optimal trading
strategy is known as being a band policy: it contains a
continuous and bounded no-trading zone, and the strategy
instantaneously trades towards this zone when being outside of
it. The challenge then is to find the exact values for the frontiers
of the no-trading zone.

A first solution to this exact problem was given in \cite{meanReversion},
however the formulation of the solution makes it very difficult to
track, except in the case of small linear costs.  In the present paper,
using a method first introduced in~\cite{optimalTh}, which infers the
limit of a no-trading zone by studying the possible future paths of
the predictor when starting from this limit, we end up with a much
more explicit solution for the upper and lower edges of the band. This allows in particular to derive new asymptotic results, which do not require costs to be small. In particular, we derive: i) the asymmetry of the band when the predictor becomes large ii) the asymptotic size of the band and iii) the position of the band around zero when trading costs become large.



\medskip

The content of the paper is as follows: after having formalized the
problem we want to solve, we show why the shape of its optimal
solution is necessarily a band (as we are not aware of any such proof
already existing in the literature for this exact problem), and then extend the techniques
introduced in~\cite{optimalTh} to derive
path-integral equations for the upper and lower edges of the band. We then
restrict ourselves to the case of a predictor following an
Ornstein-Uhlenbeck dynamics and obtain explicit solutions in this
case, for which we can derive the asymptotic behavior as a function of
the predictor's value. Finally, we run numerical estimations of our analytical formulas
and compare the resulting policy against a system with a constant and symmetric band.


\section{Description of the problem}

The problem we address in this paper is to find the optimal strategy
for a trader in the presence of a predictor, a quadratic risk penalty and a linear cost term.
This means we want to find at any moment the optimal position $\pi_t$, given:
\bi
\item A predictor of the future price returns, following a random process $(p_t)_t$, which generates a gain $p_t\cdot\pi_{t}$.
\item A risk penalty for holding a position: $\lambda\pi_t^2$.
  \item A cost penalty for trading: $\Gamma|\pi_t-\pi_{t-1}|$.
\ei


We require the predictor to be a Markovian process, independent of time $t$, and unbounded: $$\forall q\forall p,\ \exists
  \epsilon_{q,p}>0\textrm{ s.t. }\pproba(p_{t+1}>q|p_t=p)\ >\epsilon_{q,p}$$

The optimal policy can then be defined explicitely as the function $\pi^\star(\pi, p)$ given by:
\begin{equation*}
  \underset{\pi^\star:\mathbb{R}^2\rightarrow\mathbb{R}}{\textrm{argmax}}\ \ \lim_{T\rightarrow\infty}\mathbb{E}\left[\ \frac{1}{T}\sum_{t=1}^T\ p_t\pi_t-\lambda\pi_t^2-\Gamma|\pi_t-\pi_{t-1}|\
    \condition \pi_{0}=0\textrm{ and }\pi_t=\pi^\star(\pi_{t-1},p_t)\ \forall t>0\right]
  \end{equation*}

Note that without loss of generality we can rescale all the positions
by a constant factor, so we will fix the value $\lambda=1/2$. This
allows to see the value $p$ of the predictor itself as a position: it is the
position which maximizes the instantaneous gain
$g_p(\pi)=p\cdot\pi-\frac{1}{2}\pi^2$, sometimes called the
\emph{ideal position}.

Finally, we will frequently use the function $V(\pi,p)$ to indicate the future gains and losses
if we choose to stay in position $\pi$ for a value $p$ of the predictor (and then trade optimally):
\begin{gather*}
  V(\pi,p)=\mathbb{E}\left[\ \sum_{t=1}^T\ p_t\pi_t-\lambdaHalf\pi_t^2-\Gamma|\pi_t-\pi_{t-1}|\ \condition p_1=p\ ,\ \pi_0=\pi_1=\pi
    \textrm{ and }\pi_t=\pi^\star(\pi_{t-1},p_t)\ \forall t>1\ \right]
\end{gather*}
In theory $V$ should be indexed by $T$, but in practice we will
assume this $T$ to be large enough so that it does not really
intervene in the results. We have then, for any $\pi$ and $p$:
\begin{equation*}
  \pi^\star(\pi, p)=\underset{\pi'}{\textrm{argmax}}\ \left[\ V(\pi', p) - \Gamma|\pi'-\pi|\ \right]
  \end{equation*}

By expansion of its first term, $V$ also satisfies the equation:
\begin{gather*}
  V(\pi,p)=p\pi-\lambdaHalf\pi^2+\int \left[\ V(\pi^\star(\pi, p'), p')-\Gamma|\pi^\star(\pi, p')-\pi|\ \right]\proba(p'|p)\textrm{d}p'
\end{gather*}
with $\proba(p'|p)=\proba(p_{t+1}=p'|p_t=p)$.

\section{Why the band policy is optimal}

\begin{figure}[h!]
  \bec\includegraphics[width=7cm]{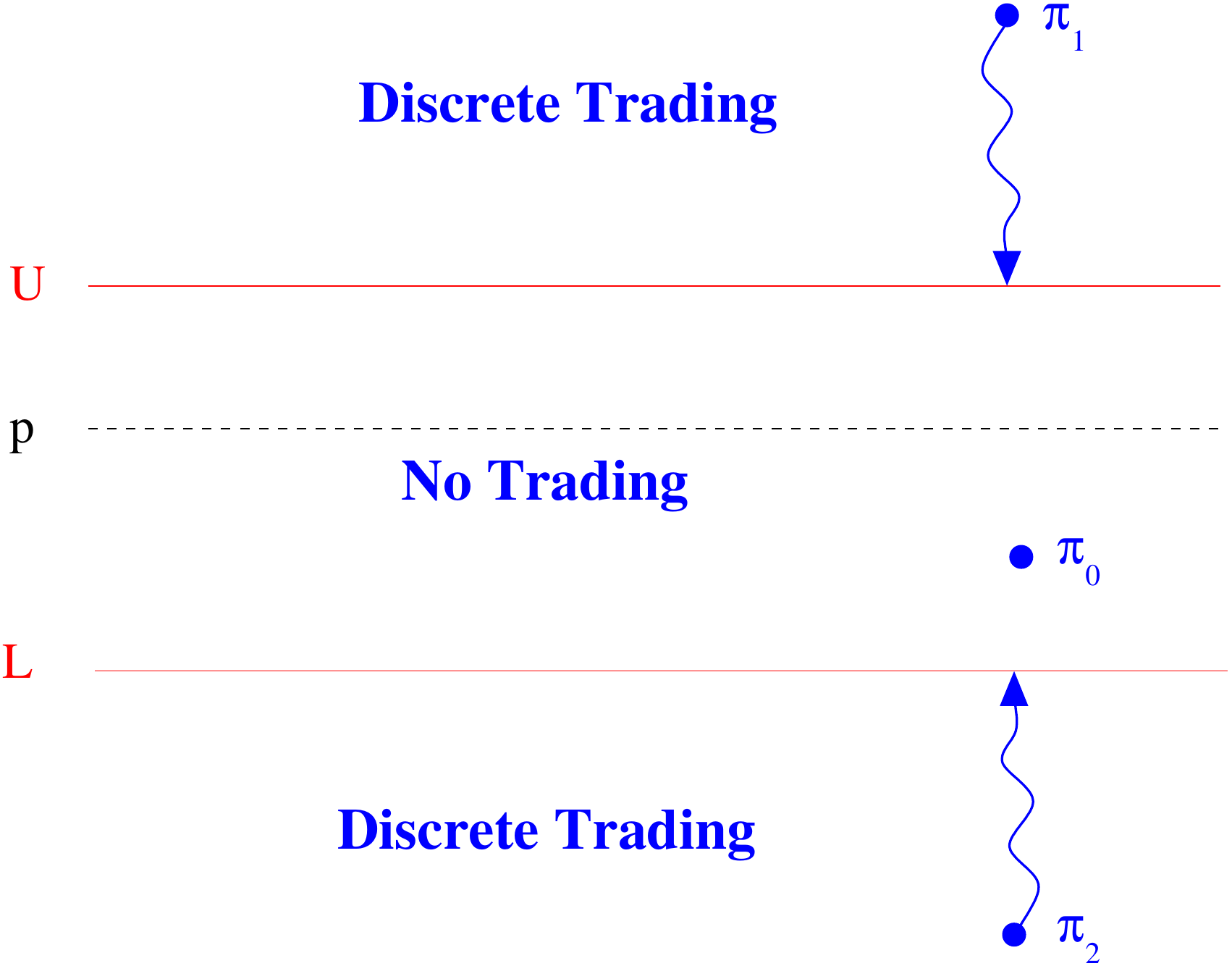}\eec
  \caption{\label{bandfigure}Behavior of the band strategy (aka. DT-NT-DT).}
\end{figure}

It is well-known folklore in the literature~\cite{johannesReview} that the
optimal strategy in this context will be a \textbf{band}, also known as
a DT-NT-DT (Discrete-Trading / No-Trading / Discrete-Trading)
policy: it is the system described on Figure~\ref{bandfigure}:

\bi
\item To each value $p$ of the predictor are associated two positions
  $\lowerBound(p)$ and $\upperBound(p)$, such that $\lowerBound(p)\leq
  p\leq\upperBound(p)$: these two positions determine a ``band''
  around the predictor.
\item If the current position is inside the band for the current predictor $p_t$, the optimal policy is to do nothing: $\pi_{t}=\pi_{t-1}$.
\item If the current position is above (resp. below) the band, the optimal policy is to trade directly towards it: $\pi_{t}=\upperBound(p_t)$ (resp. $\pi_{t}=\lowerBound(p_t)$).
\ei

This policy is highly sparse on trades, which is coherent with the $L^1$ constraint of the cost penalty.
However we are not aware of any formal justification in the literature for the optimality of that system, so in this section we would like to provide some arguments in that direction.
The proof will be made in four parts:
\begin{enumerate}
\item The function $V$ is concave in $\pi$: $\frac{\partial^2 V}{\partial\pi^2}<0$.
\item For a given $p$, the no-trading zone $\{\pi\in\mathbb{R}|\pi^\star(\pi, p)=\pi\}$ is convex, so it is a segment.
\item When we are outside of the no-trading zone, we always trade towards the edge of it.
\item The predictor $p$ is always inside the no-trading zone.
\end{enumerate}

\subsection{The function $V$ is concave in $\pi$}

Let us consider a given position $\pi$ and a fixed $p$.  Setting
current time at zero, we consider a path $(p_t)_{t\geq 0}$ for the
future evolution of the predictor, and we call $\tau$ the first time
in the future where $\pi^\star(\pi, p_\tau)=\pi_1\neq\pi$: $\tau$
will be the first moment where we do a trade.

If we call $\delta V$ the component of $V(\pi,p)$ coming from this particular future path, we have:
$$\delta V=\sum_{t=0}^\tau(p_t\pi-\lambdaHalf
\pi^2)-\Gamma|\pi_1-\pi|+V(\pi_1, p_\tau)$$
so that:
\begin{align*}
  \frac{\partial^2 \delta V}{\partial\pi^2}&=\frac{\partial^2}{\partial\pi^2}\sum_{t=0}^\tau(p_t\pi-\lambdaHalf\pi^2)\\
  &=-\tau
\end{align*}
By summing over all possible
future paths, we obtain that the second derivative along $\pi$ is indeed
negative.


\subsection{The no-trading zone is a segment}

Consider three positions $\pi_1<\pi_2<\pi_3$ for a given $p$,
suppose that $\pi_1,\pi_3$ are in the no-trading zone whereas $\pi_2$
is not. Then $\pi^\star(\pi_2, p)=\pi_4=\pi_2+\delta\pi$ with $\delta\pi\neq 0$.

Suppose that $\delta\pi>0$. Then
$V(\pi_4,p)-V(\pi_2,p)>\Gamma\cdot\delta\pi$. By the mean value theorem
there exists $\pi_5\in[\pi_2,\pi_4]$ such that $$\frac{\partial V}{\partial\pi}(\pi_5,p)=\frac{V(\pi_4,p)-V(\pi_2,p)}{\delta\pi}$$
Since $\frac{\partial^2 V}{\partial\pi^2}<0$ everywhere, we would have $\frac{\partial V}{\partial\pi}(\pi_1,p)>\Gamma$.
So, close enough around $\pi_1$, it would be worth trading: $\pi_1$ could not belong to the non-trading zone.

Of course we can apply the same argument if $\delta\pi<0$ by using
$\pi_3$ instead of $\pi_1$. So for any $p$, the no-trading zone is a
convex set on $\mathbb{R}$, hence a segment $\left[\lowerBound(p), \upperBound(p)\right]$.

\subsection{When outside the band, one trades towards its edge}

First, we prove that if
$\pi^\star(\pi_1, p)=\pi_2$ then $\pi^\star(\pi_2, p)=\pi_2$: after a
trade, we always end up in the no-trading zone. Indeed, if we had
$\pi^\star(\pi_2, p)=\pi_3\neq\pi_2$ then we would have:
\begin{align*}
  V(\pi_3,p)-V(\pi_1,p)&=V(\pi_3,p)-V(\pi_2,p)+V(\pi_2,p)-V(\pi_1,p)\\
  &>\Gamma|\pi_3-\pi_2|+\Gamma|\pi_2-\pi_1|\\
  &>\Gamma|\pi_3-\pi_1|
\end{align*}
so that, starting from $\pi_1$, it would be better to jump to $\pi_3$ than to $\pi_2$.

Moreover, this trade is always towards the edge of the band: indeed, if we have
$\pi_1<\pi_2<\pi_3$ and $\pi^\star(\pi_1, p)=\pi_3$ then
$$V(\pi_3,p)-V(\pi_1,p)-\Gamma|\pi_3-\pi_1|>V(\pi_2,p)-V(\pi_1,p)-\Gamma|\pi_2-\pi_1|$$
(otherwise we would better jump to $\pi_2$ than $\pi_3$), so:
\begin{align*}
  V(\pi_3,p)-V(\pi_2,p)&>\Gamma|\pi_3-\pi_1|-\Gamma|\pi_2-\pi_1|\\
  &>\Gamma|\pi_3-\pi_2|
\end{align*}
so $\pi_2$ is not in the no-trading zone.

\subsection{The ideal position is inside the band: $p\in[\lowerBound(p), \upperBound(p)]$}

The position $p$ is the maximum of the function $g_p(\pi)=p\cdot\pi-\frac{1}{2}\pi^2$.
By definition of $V$, for any $\pi$ we have:
$$V(\pi,p)=g_p(\pi)+\int\left(V(\pi_{p'}, p') -\Gamma|\pi_{p'}-\pi|\right)\proba(p'|p)\textrm{d}p'$$
with $\pi_{p'}=\pi^\star(\pi, p')$, so
\begin{align*}
  V(\pi, p)-\Gamma|\pi-p|&\leq g_p(p)+\int\left(V(\pi_{p'},p') -\Gamma|\pi_{p'}-\pi|-\Gamma|\pi-p|\right)\proba(p'|p)\textrm{d}p'\\
  &\leq g_p(p)+\int\left(V(\pi_{p'}, p') -\Gamma|\pi_{p'}-p|\right)\proba(p'|p)\textrm{d}p'\\
  &\leq V(p, p)
\end{align*}
so that $\pi^\star(p, p)=p$: the predictor is always inside the no-trading zone.

\bigskip

Now that we have established the shape of the optimal strategy, we
will derive the explicit equations for the values of $\upperBound(p)$
and $\lowerBound(p)$.  As already said, some equations of this sort already appear
in~\cite{meanReversion}, but here we will provide more explicit solutions that
will allow to calculate in Section~\ref{asymptotic} the asymptotic
behavior in $p$.

\section{Equations for the edges of the band}

As in~\cite{optimalTh}, we will rely on an analysis of the optimal behavior when the position is close to the non-trading zone in order to establish the equations for the band.
However, since this time we have two parameters to determine (the two edges of the band), we need to find a system of two equations.

Let us consider a value $p_1$ for the predictor, we will note $\upperBound=\upperBound(p_1)$ and $\lowerBound=\lowerBound(p_1)$.
We also introduce $p_2$ as \textbf{the value of the predictor for which $\lowerBound$ is the upper edge}: $\upperBound(p_2)=\lowerBound(p_1)$.

We suppose that the current position ($t=0$) is at $\lowerBound$, and consider two cases:
\begin{enumerate}[i)]
\item\label{predInp1} The current value of the predictor is $\underline{p_1}$, and we wonder if it is worth \underline{buying} an infinitesimal quantity $\delta\pi$.
  \item\label{predInp2} The current value of the predictor is $\underline{p_2}$ and we wonder if it is worth \underline{selling} an infinitesimal quantity $\delta\pi$.
\end{enumerate}

\begin{figure}
\bec\includegraphics[height=4.3cm]{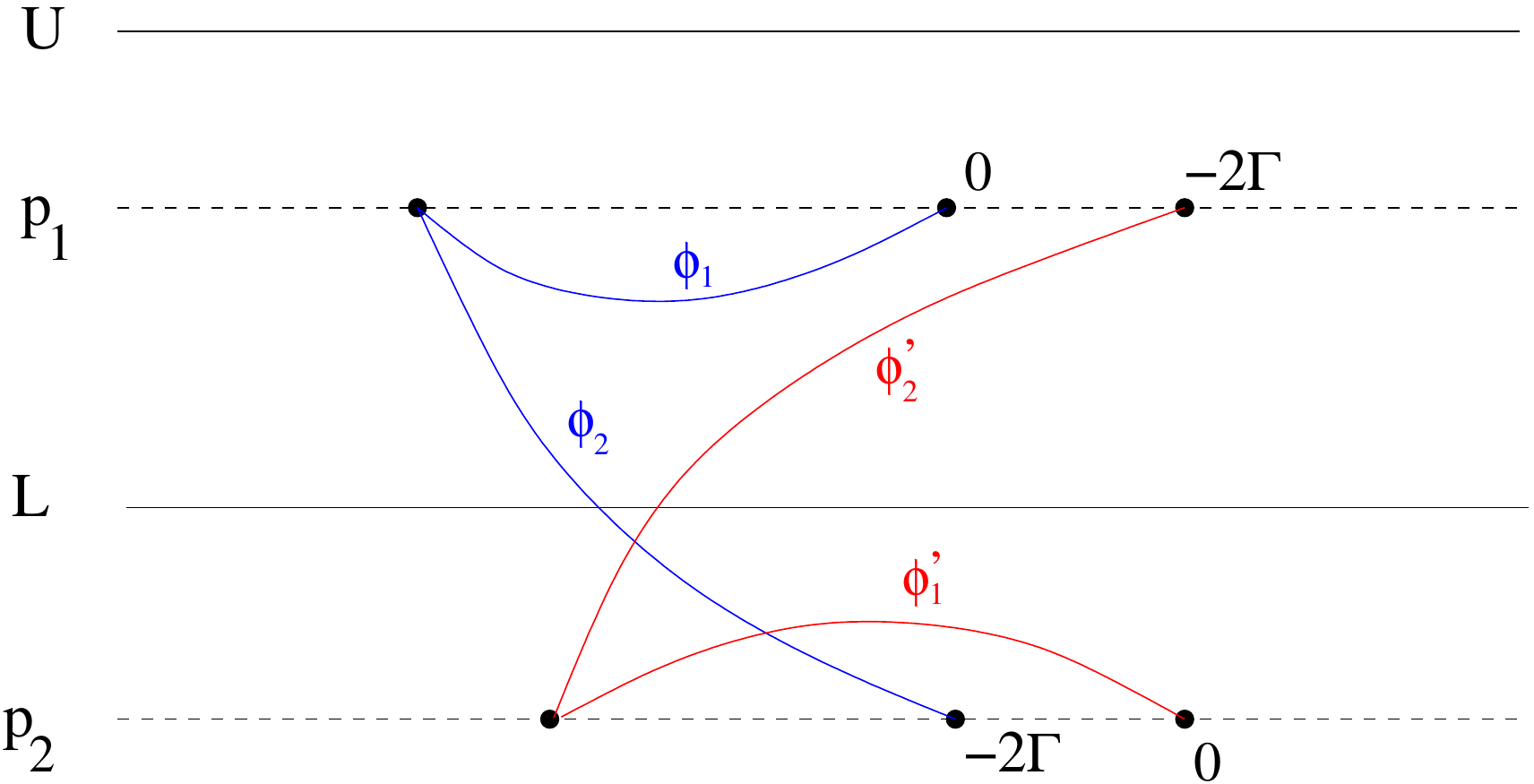}\eec
\caption{\label{pathIntegral}Configurations giving rise to the equations for the band.}
\end{figure}

In each case we will consider the different future paths taken by the
predictor, keeping in mind that our future behaviour is the optimal
one (stay inside the band or trade towards it).  The situation is
summarised on Figure~\ref{pathIntegral}. Note that we do not need to
look at what happens after we exit the band, because the optimal
position will not depend anymore on what we did at $t=0$. Note also that,
because the predictor dynamics is unbounded, the paths that stay
inside the band forever have a null contribution when we integrate over all paths, so we can safely ignore them.

\bigskip

Let us consider first the case~\ref{predInp1}). If we buy $\delta\pi$ starting from position $\lowerBound$ then we are inside the band, and we will stay there as long as:
\bi
\item either the predictor becomes larger than $p_1$ (path $\phi_1$),
  \item or it becomes smaller than $p_2$ (path $\phi_2$).
    \ei

    Compared to the case where we stayed at $\lowerBound$ without buying, we will not have suffered any additional cost if the predictor follow the path $\phi_1$, whereas we will have paid
    $2\Gamma\cdot \delta\pi$ in the case of the path $\phi_2$ (because we paid linear costs when buying $\delta\pi$, and then again by selling it when the predictor goes below $p_2$).
    We denote by $\delta\pathCost$ this potential additional cost.

    Now, in terms of gains, the difference between both situations is simply $$\delta\pathGain=\sum_{t=0}^{T_\phi}\phi(t)\cdot\delta\pi$$
    for $\phi\in\{\phi_1,\phi_2\}$, where $T_{\phi}$ is the first time where $\phi(t)>p_1$ or $\phi(t)<p_2$.

    And finally, in terms of risk, the difference is
    \begin{align*}\delta\pathRisk&=\sum_{t=0}^{T_\phi}\left(\lambdaHalf\cdot(\lowerBound+\delta\pi)^2-\lambdaHalf\cdot\lowerBound^2\right)\\
      &=T_\phi\cdot\lowerBound\cdot\delta\pi+\mathcal{O}(\delta\pi^2)
    \end{align*}
    for $\phi\in\{\phi_1,\phi_2\}$.

    We now need to integrate over all possible paths: for a finite path $\phi:[0,n]\rightarrow\mathbb{R}$, we note:
    \begin{eqnarray*}
    T_\phi&=&n\\
    \phi_b&=&\phi(0)\\
    \phi_e&=&\phi(n)\\
    \proba(\phi|p)&=&\pproba(p_{z}=\phi(z),\ z\in[0,n]\ |\ p_0=p)\\
    \int_zF(\phi(z))\textrm{d}z&=&\sum_{i=0}^{n-1}F(\phi(i))
  \end{eqnarray*}
  
    Then it is worth buying $\delta\pi$ at $t=0$ if, and only if:

    $$\int\limits_{\substack{\phi_b=p_1\\ p_2<\phi(z)<p_1,\
    z\in]0,T_\phi[}}^{\phi_e\geq p_1\ \vee\ \phi_e\leq p_2}\ \left[\ \delta\pathGain-\delta\pathRisk-\delta\pathCost\ \right] \pproba(\phi|p_1)\ \pathInt\phi\geq 0$$

By substituting with the values above, this leads to:
    
$$\hspace{-0.6cm}\delta\pi\cdot\int\limits_{\substack{\phi_b=p_1\\ p_2<\phi(z)<p_1,\
    z\in]0,T_\phi[}}^{\phi_e\geq p_1\ \vee\ \phi_e\leq p_2}\ \left[\ \int_z(\phi(z)-\lowerBound)\
  \textrm{d}z -2\Gamma\cdot\mathbf{1}_{\{\phi_e\leq p_2\}}(\phi)\
\right]\ \pproba(\phi|p_1)\ \pathInt\phi\ \geq 0$$
with $\mathbf{1}$ the indicator function.

The optimal band is such that the lower edge $\lowerBound$ is the exact position where this marginal gain is exactly zero, so we obtain our first equation:
\begin{equation}\label{firstEq}\hspace{-0.6cm}\boxed{\qquad\int\limits_{\substack{\phi_b=p_1\\ p_2<\phi(z)<p_1,\
    z\in]0,T_\phi[}}^{\phi_e\geq p_1\ \vee\ \phi_e\leq p_2}\ \left[\ \int_z(\phi(z)-\lowerBound)\
  \textrm{d}z -2\Gamma\cdot\mathbf{1}_{\{\phi_e\leq p_2\}}(\phi)\
  \right]\ \pproba(\phi|p_1)\ \pathInt\phi\ =0\qquad}
  \end{equation}

This equation is very similar to the one found in~\cite{optimalTh}, with the addition of the risk component through the term $-\lowerBound$.

\bigskip

Now we can consider case~\ref{predInp2}), where the predictor starts at $p_2$.
If we sell $\delta\pi$ starting from position $\lowerBound$ then we are inside the band, and we will stay there as long as:
\bi
\item either the predictor becomes smaller than $p_2$ (path $\phi'_1$),
  \item or it becomes larger than $p_1$ (path $\phi'_2$).
    \ei

    Compared to the case where we stayed at $\lowerBound$ without selling, we have:
    \begin{enumerate}
  \item An extra cost $\delta\pathCost=2\Gamma|\delta\pi|$ only in the cases where we the predictor becomes eventually larger than $p_1$.
  \item A difference in gain equal to:
    $$\delta\pathGain=-\sum_{t=0}^{T_\phi}\phi(t)\cdot\delta\pi$$
  \item A difference in risk equal to:
    \begin{align*}\delta\pathRisk&=\sum_{t=0}^{T_\phi}\left(\lambdaHalf\cdot(\lowerBound-\delta\pi)^2-\lambdaHalf\cdot\lowerBound^2\right)\\
      &=-T_\phi\cdot\lowerBound\cdot\delta\pi+\mathcal{O}(\delta\pi^2)
    \end{align*}
    \end{enumerate}

    So it is indeed worth selling $\delta\pi$ if, and only if:
    $$\hspace{-0.6cm}\delta\pi\cdot\int\limits_{\substack{\phi_b=p_2\\ p_2<\phi(z)<p_1,\
    z\in]0,T_\phi[}}^{\phi_e\geq p_1\ \vee\ \phi_e\leq p_2}\ \left[\ \int_z(-\phi(z)+\lowerBound)\
  \textrm{d}z -2\Gamma\cdot\mathbf{1}_{\{\phi_e\geq p_1\}}(\phi)\
\right]\ \pproba(\phi|p_2)\ \pathInt\phi\ \geq 0$$

The now upper edge $\lowerBound$ is the exact position where this marginal gain is exactly zero, so we obtain the second equation:

\begin{equation}\label{secondEq}\hspace{-0.6cm}\boxed{\qquad\int\limits_{\substack{\phi_b=p_2\\ p_2<\phi(z)<p_1,\
    z\in]0,T_\phi[}}^{\phi_e\geq p_1\ \vee\ \phi_e\leq p_2}\ \left[\ \int_z(\phi(z)-\lowerBound)\
  \textrm{d}z +2\Gamma\cdot\mathbf{1}_{\{\phi_e\geq p_1\}}(\phi)\
\right]\ \pproba(\phi|p_2)\ \pathInt\phi\ = 0\qquad}\end{equation}

\bigskip
\bigskip


For what comes next it will be useful to decompose Equations~(\ref{firstEq}) and~(\ref{secondEq}), so we set:
\begin{align*}\pathGain(p)&=
\int\limits_{\substack{\phi_b=p\\ p_2<\phi(z)<p_1,\
    z\in]0,T_\phi[}}^{\phi_e\geq p_1\ \vee\ \phi_e\leq p_2}\ \left[\ \int_z\phi(z)\ \textrm{d}z\ \right]
    \ \pproba(\phi|p)\ \pathInt\phi\\
&\\
\pathRisk(p)&=
\int\limits_{\substack{\phi_b=p\\ p_2<\phi(z)<p_1,\
    z\in]0,T_\phi[}}^{\phi_e\geq p_1\ \vee\ \phi_e\leq p_2}\ \left[\ \int_z\ \textrm{d}z\ \right]
    \ \pproba(\phi|p)\ \pathInt\phi\\
&\\
\pathCostP(p)&=
\int\limits_{\substack{\phi_b=p\\ p_2<\phi(z)<p_1,\
    z\in]0,T_\phi[}}^{\phi_e\leq p_2}\ \pproba(\phi|p)\ \pathInt\phi\\
\end{align*}
and the equations become:
\begin{align*}
\pathGain(p_1)-\lowerBound\cdot\pathRisk(p_1)-2\Gamma\cdot\pathCostP(p_1)&=0\\
\pathGain(p_2)-\lowerBound\cdot\pathRisk(p_2)-2\Gamma\cdot\pathCostP(p_2)&=-2\Gamma
\end{align*}

In the next section we will consider a continuous dynamics for the predictor, in which case each term in the above equations is equal to zero by definition (except $\pathCostP(p_2)$ which goes to $1$), and the equations become trivial. This is the classical issue of evaluating a continuous stochastic system close to a boundary, and this is solved by requiring the equalities above to be true around $p_1$ and $p_2$ up to first-order expansion\footnote{One can understand this by considering only one discrete, infinitesimal step starting from $p_1$ or $p_2$, followed by a continuous dynamics.}:
  \begin{align}
  \pathGain'(p_1)-\lowerBound\cdot\pathRisk'(p_1)-2\Gamma\pathCostP'(p_1)=0\label{equationGRC1}\\
  \pathGain'(p_2)-\lowerBound\cdot\pathRisk'(p_2)-2\Gamma\pathCostP'(p_2)=0\label{equationGRC2}
  \end{align}


\section{Case of an Ornstein-Uhlenbeck predictor}

Let us now consider the case where the dynamics of the predictor $(p_t)_t$ is given by a discrete Ornstein-Uhlenbeck process:
\begin{equation}\label{dynamicsOU}
p_{t+1}-p_t=-\varepsilon\cdot p_t+\beta\cdot \xi_t
\end{equation}
where $(\xi_t)_{t\in\mathbb{R}}$ is a set of independent $\mc{N}(0, 1)$ Gaussian random variables.

In what follows, contrary to~\cite{optimalTh}, we will only consider the continuous limit: $\beta\ll\Gamma$ (no single-step jump in the predictor is significant compared to the costs).
The dynamics of the predictor can then be written in a more continuous form:
\begin{equation}
\label{driftDiffusion}\textrm{d}p=-\varepsilon p\ \textrm{d}t + \beta\ \textrm{d}X_t
\end{equation}
where $(X_t)_t$ is a Wiener process.

\subsection{Explicit solutions}


Now that the dynamics of the predictor is fixed, we can calculate the functions $\pathGain$, $\pathRisk$ and $\pathCostP$, and solve Equations~(\ref{equationGRC1}) and~(\ref{equationGRC2}).
To make the reasonings easier to follow, we will redefine them temporarily as functions of two variables: $\pathGain(p,t)=\pathGain(p)$, $\pathRisk(p,t)=\pathRisk(p)$ and $\pathCostP(p,t)=\pathCostP(p)$.

\bigskip

To calculate $\pathGain$, we can make use of It\={o}'s 
lemma with Equation~(\ref{driftDiffusion}):
\begin{align*}\textrm{d}\pathGain&\ =\ \frac{\partial\pathGain}{\partial t}\ \textrm{d}t\ +\ \frac{\partial\pathGain}{\partial p}\ \textrm{d}p\ +\ \frac{1}{2}\beta^2\ \frac{\partial^2\pathGain}{\partial p^2}\ \textrm{d}t\\
\textrm{d}\pathGain&\ =\ \left(\frac{\partial\pathGain}{\partial t}\ -\ \varepsilon p\ \frac{\partial\pathGain}{\partial p}\ +\ \frac{1}{2}\beta^2\ \frac{\partial^2\pathGain}{\partial p^2}\right)\textrm{d}t
\ +\ \beta\ \frac{\partial\pathGain}{\partial p}\ \textrm{d}X_t\end{align*}

Let us now consider the operator $\langle\cdot\rangle_{\textrm{d}X}$ which integrates over all possible values for $\textrm{d}X_t$: by definition of
$\pathGain$, we can write, for $p\in[p_2,p_1]$, $$\pathGain(p,t)=p\ \textrm{d}t+\langle\ \pathGain(p+\textrm{d}p,t+\textrm{d}t)\ \rangle_{\textrm{d}X}$$
so that $\langle\ \textrm{d}\pathGain\ \rangle_{\textrm{d}X} = -p\ \textrm{d}t$.
Since we also have $\langle\ \textrm{d}X_t\ \rangle_{\textrm{d}X} = 0$ and $\partial\pathGain\ /\ \partial t=0$,
it gives
\begin{equation*}
\frac{1}{2}\beta^2\ \frac{\partial^2\pathGain}{\partial p^2}\ -\ \varepsilon p\ \frac{\partial\pathGain}{\partial p}\ =\ -p
\end{equation*}
with two initial conditions $\pathGain(p_1)=0$ and $\pathGain(p_2)=0$.
This is the \textbf{Kolmogorov backward equation} of the system for the gain term.

This equation can be solved as:
$$\pathGain(p)=\frac{1}{\varepsilon}\left(p-p_1-\frac{p_2-p_1}{I}\int_{p_1}^{p}e^{ax^2}\dd x\right)$$
with $$a=\frac{\varepsilon}{\beta^2}\textrm{\quad and\quad}I=\int_{p_1}^{p_2}e^{ax^2}\dd x$$

\bigskip

A similar reasoning can be applied to find the Kolmogorov backward equation for $\pathRisk$:
\begin{equation*}
  \frac{1}{2}\beta^2\ \frac{\partial^2\pathRisk}{\partial p^2}\ -\ \varepsilon p\ \frac{\partial\pathRisk}{\partial p}\ =\ -1
  \end{equation*}
with initial conditions $\pathRisk(p_1)=\pathRisk(p_2)=0$.

Its solution is:
$$\pathRisk(p)=\frac{2aK}{\varepsilon}\left(\frac{1}{I}\int_{p_1}^{p}e^{ax^2}\dd x
-\frac{1}{K}\int_{p_1}^{p}e^{ax^2}\left[\int_{p_1}^{x}e^{-ay^2}\dd y\right]\dd x\right)$$
with
$$K=\iint\limits_{p_2\leqslant x\leqslant y\leqslant p_1}e^{a(x^2-y^2)}\ \dd x\ \dd y$$

\bigskip

And finally, the equation for $\pathCostP$ is:
\begin{equation*}
  \frac{1}{2}\beta^2\ \frac{\partial^2\pathCostP}{\partial p^2}\ -\ \varepsilon p\ \frac{\partial\pathCostP}{\partial p}\ =\ 0
  \end{equation*}
with initial conditions $\pathCostP(p_1)=0$ and $\pathCostP(p_2)=1$.

Its solution is:
$$\pathCostP(p)=\frac{1}{I}\int_{p_1}^{p}e^{ax^2}\dd x$$

\bigskip


Plugging the functions above into Equations ~(\ref{equationGRC1}) and~(\ref{equationGRC2}) (with unknown $p_2$ and $\ell$), we obtain:
$$\begin{cases}
  Ie^{-ap_1^2}-(p_2-p_1)-2a\ell\cdot K-2\Gamma\varepsilon&=0\\
  Ie^{-ap_2^2}-(p_2-p_1)-2a\ell\cdot K+2a\ell\cdot IJ-2\Gamma\varepsilon&=0\\
  \end{cases}$$
with $$J=\int_{p_1}^{p_2}e^{-ax^2}\dd x$$

By simply solving this system of two equations, we finally end up with the result:

\begin{Proposition}\label{mainResultOU}
For a predictor whose dynamics is governed by Equation~(\ref{dynamicsOU}), the lower edge of the band associated to a value $p=p_1$ of the predictor is:
\begin{equation}\label{equationLowerBound}\lowerBound(p_1)=\frac{e^{-ap_1^2}-e^{-ap_2^2}}{2a\cdot J}\end{equation}
where $p_2$ is given, as a function of $p_1$, by:
\begin{equation}\label{equation_p1_p2}
  \boxed{\quad p_1-p_2\ =\ 2\Gamma\varepsilon-Ie^{-ap_1^2}+\frac{K\cdot(e^{-ap_1^2}-e^{-ap_2^2})}{J}\quad}
\end{equation}
with:
$$a=\frac{\varepsilon}{\beta^2}\ ,\quad I=\int_{p_1}^{p_2}e^{ax^2}\dd x\ ,\quad J=\int_{p_1}^{p_2}e^{-ax^2}\dd x\ ,\quad
K=\iint\limits_{p_2\leqslant x\leqslant y\leqslant p_1}e^{a(x^2-y^2)}\ \dd x\ \dd y$$

Similarly, the upper edge of the band associated to a value $p=p_2$ of the predictor is:
\begin{equation}\label{equationUpperBound}\upperBound(p_2)=\frac{e^{-ap_1^2}-e^{-ap_2^2}}{2a\cdot J}\end{equation}
where $p_1$ is given, as a function of $p_2$, by Equation~(\ref{equation_p1_p2}).
\end{Proposition}

\bigskip
All the parameters of the problem can actually be factorized in Proposition~\ref{mainResultOU}: indeed, if we set $q_1=p_1\sqrt{a}$, $q_2=p_2\sqrt{a}$ and $\lowerBound^{\star}=\lowerBound\sqrt{a}$, then the result can be rewritten as:
\begin{equation}\label{eqNoVar}\lowerBound^{\star}(q_1)=F(q_1, q_2)\textrm{\qquad with $q_2$ given by:\qquad }G(q_1, q_2)=\frac{2\Gamma\varepsilon^{3/2}}{\beta}\end{equation}
where
\begin{align*}
  F(q_1, q_2)&=\frac{e^{-q_1^2}-e^{-q_2^2}}{2\int_{q_1}^{q_2}e^{-x^2}\dd x}\\
  G(q_1, q_2)&=q_1-q_2+e^{-q_1^2}\int_{q_1}^{q_2}e^{x^2}\dd x-\frac{e^{-q_1^2}-e^{-q_2^2}}{\int_{q_1}^{q_2}e^{-x^2}\dd x}\cdot\iint\limits_{q_2\leqslant x\leqslant y\leqslant q_1}e^{x^2-y^2}\ \dd x\ \dd y
  \end{align*}

As explained in~\cite{optimalTh}, up to a factor $\sqrt{2}$, $\beta\varepsilon^{-1/2}$ is the standard deviation $\sigma_p$ of the predictor and $\beta\varepsilon^{-3/2}$ its \emph{integrated} average gain
(taking into account its autocorrelation).
So the factor $\Gamma / \beta\varepsilon^{-3/2}$ from Equation~(\ref{eqNoVar}) is a very natural scale for the problem, since it compares the average total gain coming from the predictor to the cost of a trade. 
The rescaling $q=p\sqrt{a}=p / \beta\varepsilon^{-1/2}$ is also easy to interpret: it is just a rescaling of the predictor by its standard deviation $\sigma_p$ (multiplied by $\sqrt{2}$).

So, after normalisation of the predictor, the edges of the band are only determined by the predictor's value and the ratio $\Gamma / \beta\varepsilon^{-3/2}$.

\subsection{Asymptotic shape of the band}\label{asymptotic}

Now that we have the explicit solutions for $\lowerBound(p)$ and $\upperBound(p)$ through Proposition~\ref{mainResultOU}, we can look at their asymptotic behavior when the predictor $p$ takes very large or very small values. 


The questions we are interested in are the following:
\bi
\item How does the \textbf{size} of the band evolve with large / small values of $p$?
  \item How is the \textbf{symmetry} of the band around the predictor affected in those limits?
  \ei

\subsubsection{Case where $p\rightarrow 0$}

If $p_1=0$, the symmetry of the system is straightforward: $\upperBound(0)=-\lowerBound(0)$.
Now, using the notation $p=p_2$ for simplicity, Equation~(\ref{eqNoVar}) becomes:
$$G(0, p\sqrt{a})=\frac{2\Gamma\varepsilon^{3/2}}{\beta}$$
As we would like to consider the limit $p\rightarrow 0$, this requires $\Gamma$ to be small, more precisely: $$\Gamma\ll\beta\varepsilon^{-3/2}$$
  In this limit, one has then:
\begin{align*}
  I&=p+\frac{a}{3}\cdot p^3+\mathcal{O}(p^4)\\
  J&=p-\frac{a}{3}\cdot p^3+\mathcal{O}(p^4)\\
  K&=\frac{p^2}{2}+\mathcal{O}(p^4)
  \end{align*}
So Equation~(\ref{equation_p1_p2}) becomes, to the main order in $p$:
$$-p=2\Gamma\varepsilon-p-\frac{a}{3}\cdot p^3+\frac{\frac{p^2}{2}\cdot ap^2}{p-\frac{a}{3}\cdot p^3}$$
which leads to:
$$p^3=-12\cdot\Gamma\beta^2$$
Equation~(\ref{equationLowerBound}) then gives:
$$\lowerBound=\frac{ap^2}{2a(p-\frac{a}{3}\cdot p^3)}=\frac{p}{2}+\mathcal{O}(p^3)$$
So we obtain $$\lowerBound(0)=-\sqrt[3]{\frac{3}{2}\cdot\Gamma\beta^2}\qquad\textrm{and}\qquad\upperBound(0)=\sqrt[3]{\frac{3}{2}\cdot\Gamma\beta^2}$$

Those are the limits found in~\cite{meanReversion} as well as
in~\cite{optimalTh} in the case of small linear costs. See in
particular~\cite{delta23} where an explanation is provided for the appearance of a $1/3$
exponent on the parameter $\Gamma$.

\subsubsection{Case where $p\rightarrow +\infty$ (continuous case)}

We now consider the limit $p_1,p_2\rightarrow +\infty$. First, let us
recall that:
$$\int_0^p e^{\alpha x^2}\dd x\underset{p\rightarrow\infty}{\approxim}\frac{e^{\alpha p^2}}{2\alpha p}+C^{\textrm{te}}$$
for any $\alpha$, positive or negative. 

This leads to:
\begin{equation*}
  I\approxim-\frac{e^{ap_1^2}}{2ap_1}(1-\eta\cdot\frac{p_1}{p_2})\qquad J\approxim-\frac{e^{ap_2^2}}{2ap_2}(1-\eta\cdot\frac{p_2}{p_1})
\end{equation*}
with $$\eta=e^{a(p_2^2-p_1^2)}$$
and also: \begin{align*}
  K&\approxim\int_{p_1}^{p_2}e^{ax^2}\cdot\left(\frac{e^{-ap_1^2}}{2ap_1}-\frac{e^{-ax^2}}{2ax}\right)\dd x\\
  &\approxim\frac{e^{-ap_1^2}}{2ap_1}\int_{p_1}^{p_2}e^{ax^2}\dd x-\int_{p_1}^{p_2}\frac{\dd x}{2ax}\\
  &\approxim-\frac{1}{4a^2p_1^2}(1-\eta\cdot\frac{p_1}{p_2})-\frac{1}{2a}\ln{\frac{p_2}{p_1}}
\end{align*}

Plugging everything into Equation~(\ref{equation_p1_p2}), we obtain, to first order in $p_1$ and $p_2$:
$$p_1-p_2=2\Gamma\varepsilon+\frac{1}{2ap_1}(1-\eta\cdot\frac{p_1}{p_2})-p_2\cdot\frac{1-\eta}{1-\eta\cdot\frac{p_2}{p_1}}\left(\ln{\frac{p_2}{p_1}}+\frac{1}{2ap_1^2}(1-\eta\cdot\frac{p_1}{p_2})\right)$$
We can then assume that $p_1-p_2\ll p_1$ and $\eta\ll 1$, so we have:
$$p_1-p_2=2\Gamma\varepsilon-p_2\cdot\frac{p_2-p_1}{p_1}$$
We set $B=p_1-p_2$, to get:
$$B=2\Gamma\varepsilon+B\cdot(1-\frac{B}{p_1})$$
so:
$$B=\sqrt{2\Gamma\varepsilon\cdot p_1}$$

The equation for the lower edge of the band gives:
\begin{equation*}
  \ell\approxim p_2\cdot\frac{1-\eta}{1-\eta\cdot\frac{p_2}{p_1}}\approxim p_2
\end{equation*}

So, in this limit, the band becomes \textbf{completely asymmetric}: the upper edge is equal to the value of the predictor.
Consequently, $B=p_1-p_2$ is in fact the size of the band, which \textbf{grows as the square-root of the predictor}.

So, to summarize, the equations give:
\begin{align*}
  \upperBound(p)&\underset{p\rightarrow\infty}{\approxim} p\\
  \lowerBound(p)&\underset{p\rightarrow\infty}{\approxim} p-\sqrt{2\Gamma\varepsilon\cdot p}
  \end{align*}

\medskip

Taking a step back, the fact that the band becomes asymmetric and bigger for larger $p$ can be understood intuitively: the ideal position $\pi=p$ is
the one maximising the instantaneous gain/risk term, and the role of the band is to avoid incuring excessive costs by following this position exactly at any moment.
Now, when the predictor becomes large, it becomes extremely likely that it will revert towards zero, considering its dynamics given by Equation~(\ref{dynamicsOU}). So:
\bi
\item If we are above the ideal position, it makes sense to trade towards it since we will maximize the instantaneous gain/risk term while doing a trade that we are very likely
  to do anyway during the next time steps; hence the \emph{asymmetry} of the band.
\item If we are below the ideal position, any trade we do towards the predictor will give us an immediate reward in the gain/risk term,
  but this rewards will most likely be offset by the fact that we will have to trade back
  during the next time steps; hence the lower edge getting farer away from the predictor, and the band \emph{increasing in size} with the predictor's value.
\ei

\subsubsection{Case where $p\rightarrow +\infty$ (discrete case)}

The results above apparently imply that the size of the band will grow
indefinitely\dots\ But there is an important
pitfall there: when we introduced the continuous Ornstein-Uhlenbeck
dynamics, we stated that \textbf{no single-step jump in the predictor
  is significant compared to the costs}.

This hypothesis is in general guaranteed by the fact that
$\beta\ll\Gamma$, since $\sigma_p$ is of the order of
$\beta\epsilon^{-1/2}$. But if we take the freedom to explore very
large predictor's values for $p$, then we will reach the point where the
decrease $\varepsilon\cdot p$ coming in the next time step through
Equation~(\ref{dynamicsOU}) becomes comparable to the cost $\Gamma$. Then the
continuity hypothesis is broken, and all our calculations above are
not valid anymore.

Fortunately, in this extreme limit, the size of the band can
actually be inferred from intuitive arguments.  Suppose we are at
position $\pi_0$ slightly below the optimal lower bound, the
predictor's value $p$ being extremely large (and positive).  At the
next time step the predictor will almost certainly be below $\pi_0$,
so any trade we do in the direction of the band will have to be
reverted immediately.

For any buy trade $q>0$, one has then:
$$V(\pi_0+q, p)=p\cdot(\pi_0+q)-\frac{1}{2}(\pi_0+q)^2-2\Gamma\cdot q+\overline{V}$$
where $\overline{V}$ is independent of $q$.
So the maximum is reached when:
\begin{align*}
  \frac{\partial V}{\partial q}=0&\Leftrightarrow p-q-\pi_0-2\Gamma=0\\
  &\Leftrightarrow \pi_0+q =p-2\Gamma
\end{align*}

Since the band is totally asymmetric for this extreme value of $p$,
we obtain:
\begin{align*}
  \upperBound(p)&\underset{p\rightarrow\infty}{\approxim} p\\
  \lowerBound(p)&\underset{p\rightarrow\infty}{\approxim} p-2\Gamma
\end{align*}
so \textbf{the band size converges to $2\Gamma$}.




To summarize, when $p$ becomes large, the size of the optimal band first
grows as a square-root, as long as the system stays continuous, until we reach a region where the cost of trading are dwarfed by the instantaneous reward of the gain-risk term, and the band size then saturates. This behavior is very reminiscent of what happened in~\cite{optimalTh} to the value of the threshold when $\beta$ grows.

\section{Numerical results}

\begin{figure}[h!]
  \begin{minipage}[t]{.5\textwidth}
    \centering
    \includegraphics[width=\textwidth]{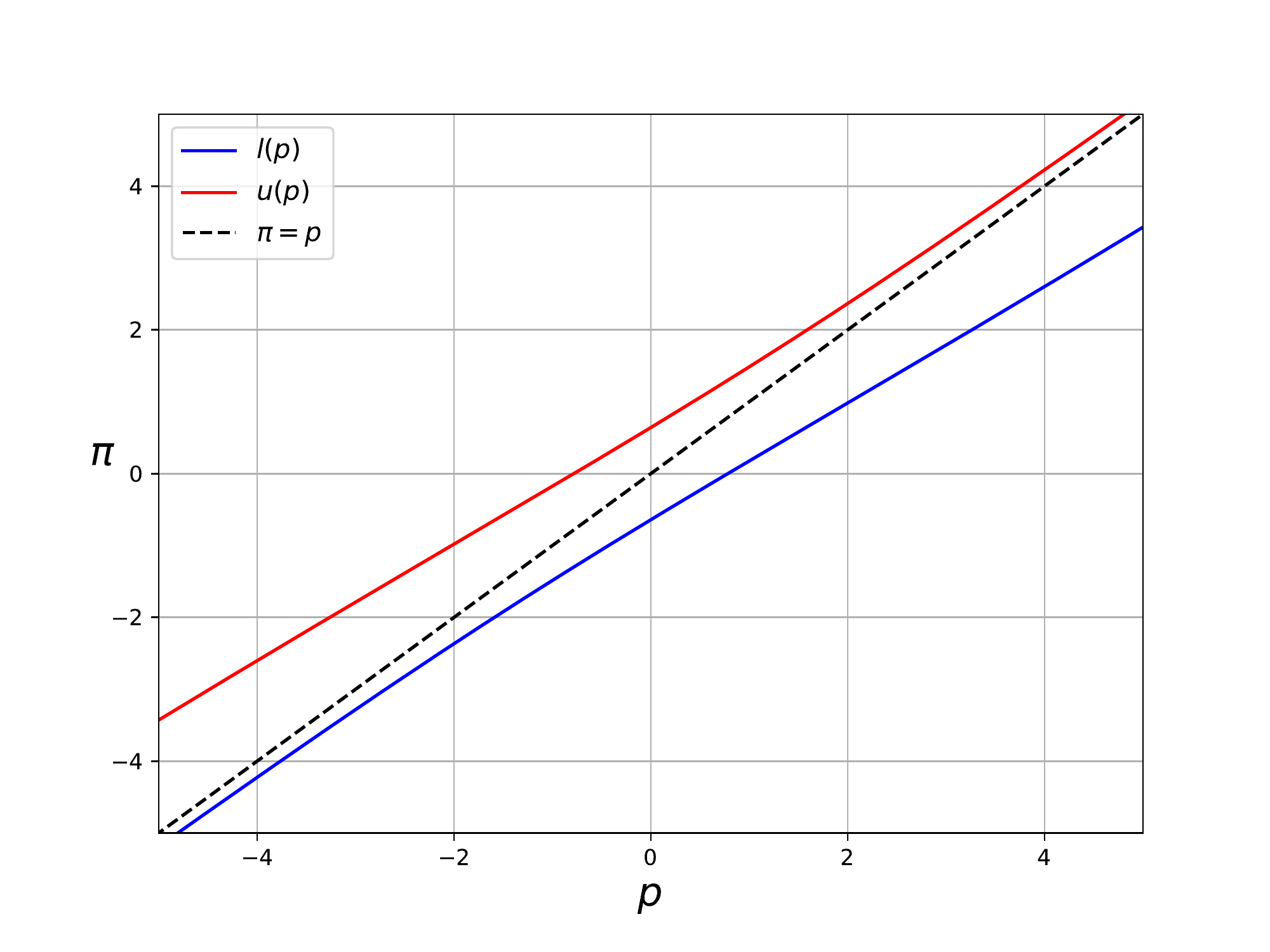}
    \subcaption{$\Gamma / \beta\varepsilon^{-3/2}=0.1$}
    \end{minipage}
  \begin{minipage}[t]{.5\textwidth}
   \centering
    \includegraphics[width=\textwidth]{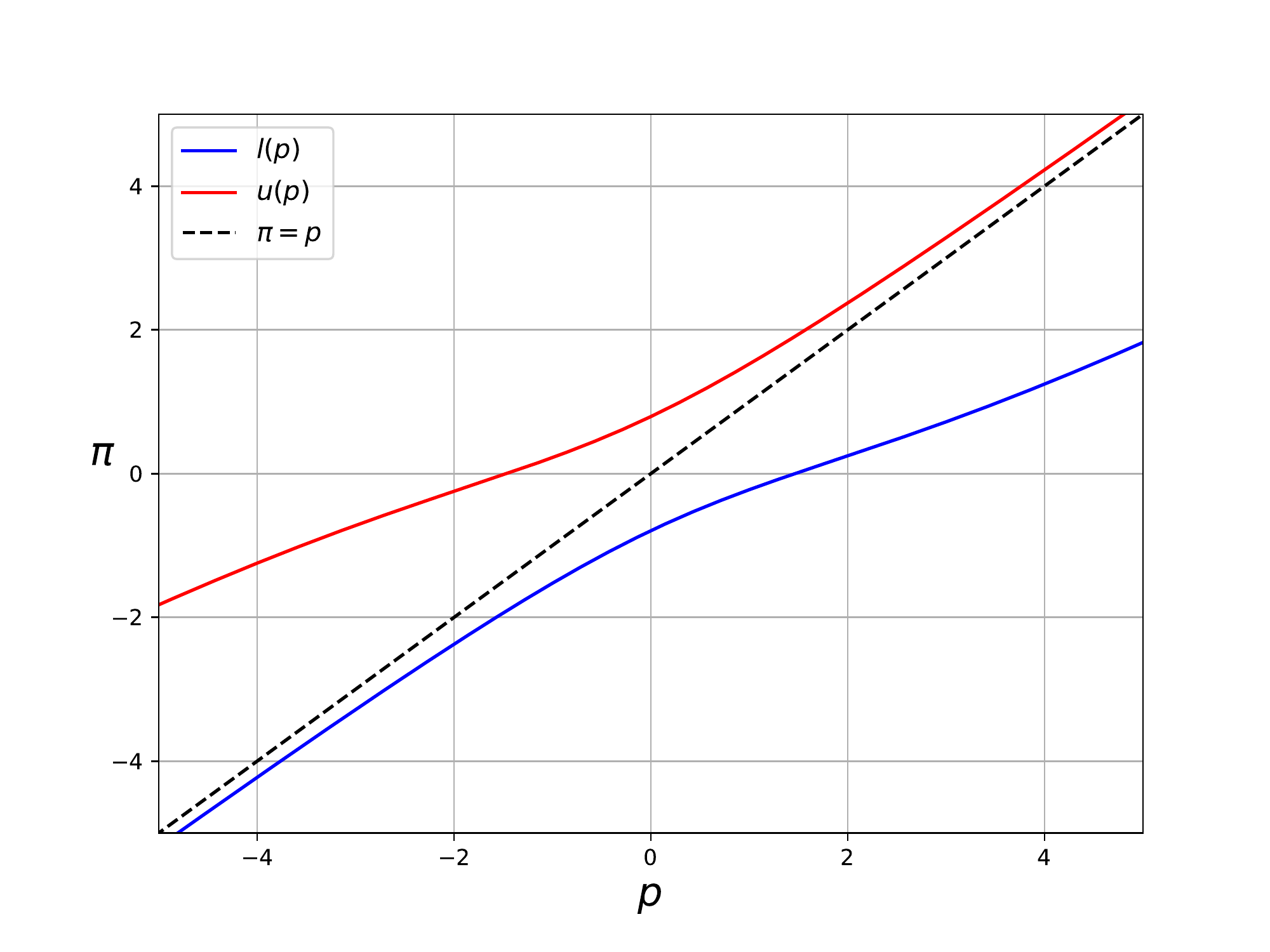}
    \subcaption{$\Gamma / \beta\varepsilon^{-3/2}=0.5$}
    \end{minipage}
    
  \begin{minipage}[t]{.5\textwidth}
    \centering
    \includegraphics[width=\textwidth]{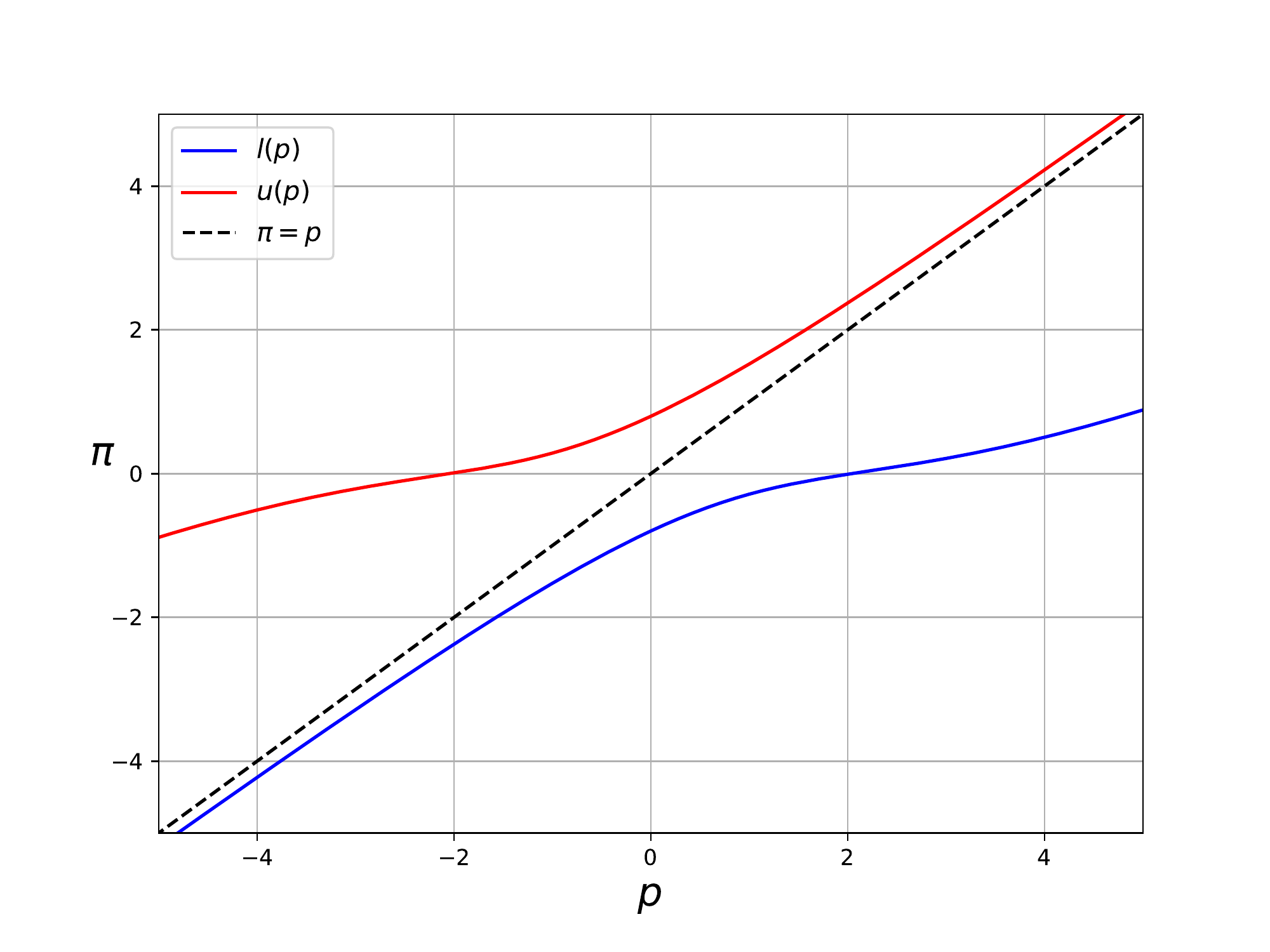}
    \subcaption{$\Gamma / \beta\varepsilon^{-3/2}=1$}
    \end{minipage}
      \begin{minipage}[t]{.5\textwidth}
    \centering
    \includegraphics[width=\textwidth]{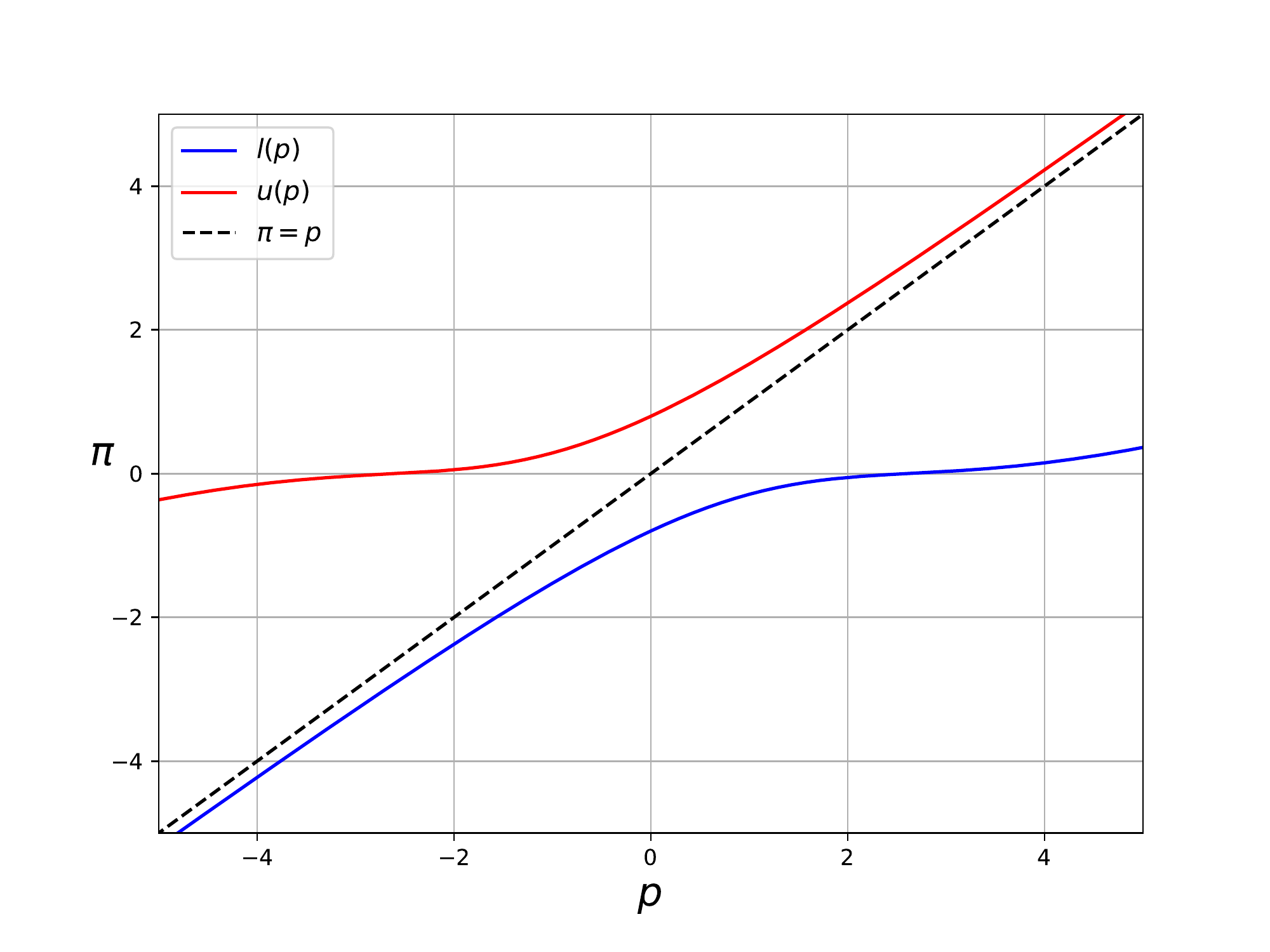}
    \subcaption{$\Gamma / \beta\varepsilon^{-3/2}=1.5$}
    \end{minipage}

  \caption{\label{numerical_results}Numerical estimations of the upper and lower edges of the band in the position vs. predictor space, for different values of $\Gamma / \beta\varepsilon^{-3/2}$. The x-axis and y-axis have been rescaled by $\sigma_p$, to make the resulting curves universal.}
\end{figure}

By inverting Equation~(\ref{equation_p1_p2}), one can find numerically the values of the lower and the upper bounds for a given value of $p$. Note that the process can be quite unstable since large exponential values are involved, so one needs
to be careful when initializing the solver. This gives in the end the results shown on Figure~\ref{numerical_results},
where the upper and lower edges of the band are shown as functions of $p$, for different values of $\Gamma$ - or, more precisely, as functions of $p/\sigma_p$ for different values of the parameter $\Gamma / \beta\varepsilon^{-3/2}$, since we want to comply with the universality of Equation~(\ref{eqNoVar}).


 \medskip

One can see several interesting results on these figures:
\begin{enumerate}
    \item The asymmetry of the band is clearly visible for all values of $\Gamma$.
    \item By contrast, the increase of the band size when $p$ grows is much more apparent for large values of $\Gamma$.
    \item The width of the (necessarily symmetric) band around $p=0$ seems to reach a maximum when $\Gamma$ grows.
\end{enumerate}

The third point in particular is interesting and rather counter-intuitive, but well supported by the equations: indeed, for large values of $\Gamma$ it is pretty clear that we will have $|p_2|\gg\sigma_p$: the value of the predictor that initiates a trade towards $\lowerBound(0)$ has to be large in order to beat the costs. So, without solving Equation~(\ref{equation_p1_p2}), we have:
\begin{equation*}
    \lim_{\Gamma\rightarrow +\infty} J=\int_0^{-\infty}e^{-ax^2}\dd x= -\frac{1}{2}\sqrt{\frac{\pi}{a}}
\end{equation*}
and consequently:
\begin{equation}\label{band_size_0}\lim_{\Gamma\rightarrow +\infty}\upperBound(0)-\lowerBound(0)=\frac{2}{\sqrt{\pi\cdot a}}= \sqrt{\frac{8}{\pi}}\cdot\sigma_p
\end{equation}

This probably deserves a little bit of explanation: why would a no-trading band reach a maximal width when linear costs become very large? The situation is in fact the following:
\bi
\item For high values of $\Gamma / \beta\varepsilon^{-3/2}$, one will have to wait for a very long time before seeing a predictor's value which justifies to trade away from $0$ (ie. which "beats its costs").
\item Consequently, when the predictor's value is zero, there is no incentive to stay in a position far from it: we will suffer a loss due to the risk term while desperately waiting for the predictor to beat its costs again. More specifically, if $\Gamma$ grows by a factor $k$, the cost of trading is multiplied by $k$, whereas the waiting time before having a value of $p$ that triggers a trade is increased exponentially, and so will be the loss due to the risk penalty.
\item However, even if we trade, the optimal policy is not to trade directly towards zero: indeed, once close enough from zero, one can afford to wait a little bit to see whether the predictor becomes positive or negative\footnote{This reasoning is interestingly reminiscent of an optimal liquidation problem with a predictor~\cite{lehallesignals}: indeed, the high value of $\Gamma$ means that one is only allowed to trade in one direction, but one can play with the value of the predictor to decide when it is best to do the trades.}. If it becomes negative (and if our position is positive), we'll have lost a little bit in risk before finishing the trade, but if it becomes positive then we can stay in position and benefit from the gain-risk term a little bit more. Furthermore, it seems like a good idea to wait at a distance of the order of the standard deviation of the predictor, since it is the order of magnitude the predictor is meant to reach in a time comparable to what it will take to come back.
\ei

So, to summarize, if the predictor is zero and the position is far from zero, then it will be brought back closer to it, but up to a point where it is comparable with the predictor's standard deviation: this is exactly what is implied by Equation~(\ref{band_size_0})~! Note that of course this limit only applies around zero, whereas the band size will continue to grow with $\Gamma$ for larger values of $p$.

\medskip

Finally, we have compared the results we obtain through the equations with a simple grid-search on a fixed and symmetric band system:
$\upperBound(p)=p+B/2$ and $\lowerBound(p)=p-B/2$, where $B$ is optimized for any tuple $\Gamma,\beta,\varepsilon$ by simply maximizing a PnL over a set of sample trajectories for the predictor.

To compare the two systems, we ran 100 simulations of 50\,000 time steps for each value of $\Gamma$ (with $\beta=\varepsilon=0.01$) and looked at the PnL after risk and cost penalties.
The results are shown on Table~\ref{comparison_constant_band}: as expected, the system induced by the equations outperforms significantly the constant and symmetric band in all cases. In particular, in the case of high linear costs when $\Gamma / \beta\varepsilon^{-3/2}=0.5$, this system is still able to generate some positive PnL whereas the more basic band avoids any trading at all.

\begin{table}[h!]
    \centering
	\begin{tabular}{c|cc|cc}
		{\quad$\Gamma / \beta\varepsilon^{-3/2}$\quad} & \multicolumn{2}{c|}{\quad Optimal Band \quad\quad} & \multicolumn{2}{c}{\quad Grid Search \quad\quad} \\
		\hline
		\hline
		0.01       & 110.44 & (0.75) & 94.13 & (0.73) \\ 
		\hline
		0.1       & 67.85 & (0.67) & 65.17 & (0.69) \\ 
		\hline
		0.15       & 54.97 & (0.63) & 49.07 & (0.67) \\ 
		\hline
		0.2       & 45.11 & (0.60) & 32.98 & (0.66) \\
		\hline
		0.3       & 30.95 & (0.53) & 17.75 & (0.55) \\ 
		\hline
		0.5       & 14.81 & (0.41) & 0 & (0)
	\end{tabular}
	    \caption{\label{comparison_constant_band}Comparing simulation results between the optimal system given by Proposition~\ref{mainResultOU} and a constant-size, symmetric band found by grid-search. Number in parentheses indicate the statistical error calculated on the sample of simulations.}
\end{table}

\section*{Conclusion}

In this paper we have given explicit solutions for the optimal edges of the band
in a system with linear costs, quadratic risk control and an
Ornstein-Uhlenbeck price predictor. This allows to study the shape of this band precisely and to derive some asymptotic behaviors of interest. Furthermore, we have shown that the 
method of analyzing paths in a no-trading zone introduced in~\cite{optimalTh} is a solid alternative to the explicit calculation of a value function, that may apply to other specific optimization problems like mixing linear
and non-linear costs~ \cite{cfmLinAndQuad} or the study of the multi-asset case~\cite{martinMulti, multi_asset}.

Another interesting direction to dig into would be to see how much of the present results can be recovered through a more exploration-based approach, using modern machine-learning methods to solve the problem. The reinforcement learning viewpoint presented in~\cite{ddpo} has been tried in the context of the present work but, for the high values of $\Gamma$ that we have been testing, we found the system to be too unstable to offer a strong benchmark against our analytical solution.




\section*{Acknowledgements}

We would like to thank Jean-Philippe Bouchaud and Stephen Hardiman for many fruitful interactions on the content of this article, as well as Johannes Muhle-Karbe for his help with academic references.

\bibliographystyle{alpha} \bibliography{biblio_band}

\end{document}